\documentclass[10pt,conference,final,letterpaper,twoside]{IEEEtran}
\usepackage{setspace}
\usepackage{amscd,graphicx,latexsym,multicol,cite,ulem,epsf}
\usepackage{verbatim}
\usepackage{graphics}
\usepackage{amssymb}
\usepackage[cmex10]{amsmath}
\IEEEoverridecommandlockouts	
\usepackage[draft,footnote,nomargin]{fixme}
\begin{document}
\title{Anticipatory Buffer Control and Resource Allocation for Wireless Video Streaming}
\author{Sanam Sadr and Stefan Valentin}

\author{Sanam Sadr\IEEEauthorrefmark{1}\IEEEauthorrefmark{2}\thanks{This work was supported by DAAD's program \emph{RISE} and by the \emph{PhD@Bell Labs} internship program.} and Stefan Valentin\IEEEauthorrefmark{2}\medskip\\
\IEEEauthorrefmark{1}Department of Electrical and Computer Engineering, University of Toronto, Canada\\
\IEEEauthorrefmark{2}Bell Labs, Alcatel-Lucent, Germany\medskip\\
{\small ssadr@comm.utoronto.ca, stefan.valentin@alcatel-lucent.com}
}
\maketitle

\begin{abstract}
This paper describes a new approach for allocating resources to video streaming traffic. Assuming that the future channel state can be predicted for a certain time, we minimize the fraction of the bandwidth consumed for smooth streaming by jointly allocating wireless channel resources and play-out buffer size. To formalize this idea, we introduce a new model to capture the dynamic of a video streaming buffer and the allocated spectrum in an optimization problem. The result is a Linear Program that allows to trade off buffer size and allocated bandwidth. Based on this tractable model, our simulation results show that anticipating poor channel states and pre-loading the buffer accordingly allows to serve more users at perfect video quality. \end{abstract}

\section{Introduction}\label{introduction}
Recent studies show that video streaming is the dominating traffic in the mobile Internet\cite{cisco}. This enormous traffic results from streaming news, entertainment clips and live events to mobile users. Video applications demand higher data rate and are delay sensitive based on the type of the video. Non-real time streams have the properties of known duration and data rate and can be buffered. The main objective is to provide a smooth video stream by ensuring that the video play-out buffer stays filled until the end of the video stream. A major challenge to guarantee smooth video streaming is the limited radio resources especially at the cell edge, or when the user is traversing an overloaded cell. 

In this paper, we consider the problem of resource allocation for users with non-real time video, e.g., as provided by the HTTP live streaming (HLS) protocol\cite{pantos11:hls_draft_ietf}. We, then, study the gain achieved in smooth video streaming based on availability of the \textit{future channel condition} and \textit{buffering}. The rational behind our proposed scheme is as follows: as the user moves away from the serving base station (BS), the channel state decreases. The channel state hence the video quality will improve if the user performs a hand-over to the adjacent cell with enough resources. However, outage\footnote{By outage, we mean there are not enough bits in the buffer to play. Hence, streaming stops until the buffer is sufficiently filled again. This is the common behavior with the HLS protocol.} will be inevitable if the next cell is overloaded. There are two ways to address this issue, if the serving base station has knowledge of the undesired future channel condition based on the video user's trajectory: (a) delaying the hand-over or (b) pre-loading the buffer. Both essentially suggest to allocate more resources from the \emph{current} BS to the video user. However, delaying the hand-over means allocating resources to the user while it is farther away from the BS resulting in higher transmit power for the same level of modulation and lower spectral efficiency. Therefore, we exploit prediction instead by pre-loading the buffer under good channel condition when a low channel gain is anticipated. Consuming the buffer then assures smooth video streaming until the improved channel state provides a stable filling rate again.  

The challenge in doing so arises from three main factors: 1) scarce resources from the serving BS; 2) the user's tendency to switch to another video without watching the pre-loaded one; 3) the user experiencing undesired channel condition or traversing an overloaded cell. The first two dictate limiting the allocated rate to the minimum video rate requirement while the third factor obligates the BS to allocate more than instantaneous video play-out bit rate to prevent outage in the near future. Due to the obvious trade-off between the mentioned factors, the technical question we attempt to solve is: \emph{how much pre-loading is enough to guarantee a smooth video?}  

The contributions of this paper are as follows: first, we propose a model for the buffer of the video user to keep track of the allocated, played and remaining bits in the buffer while meeting the video bit rate constraint. Second, we use this model to formulate an optimization problem to adjust the video user's requested rate per time slot to have enough buffered video on the one hand while consuming the minimum spectrum on the other. After summarizing the related work in Section \ref{sec:relwork}, we describe the proposed buffer control and resource allocation scheme in Section \ref{contribution}. The simulation results of the proposed scheme are presented in Section \ref{results}. The paper is concluded in Section \ref{conclusion}. 

\section{Related Work}\label{sec:relwork}
There is a large body of work on techniques for adaptive video streaming. At the application layer, channel-aware prefetching schemes invoke a traffic burst at high channel gain \cite{shih-fu:05,reisslein97:vbr_prefetching}. At the link layer, cross-layer schedulers to jointly adapt video quality and wireless resource allocation have been proposed \cite{draexler13:span,huang:08}. 

Compared to this work, our approach differs two-fold. First, it does not adapt the video quality but adjusts the size of the video play-out buffer along with the wireless data rate. Unlike \cite{draexler13:span} and \cite{huang:08}, this enables to trade off buffer size with the amount of allocated resources. Second, our adaptation scheme is \emph{anticipative}. Unlike any of the above approaches, our joint buffer-rate allocation is based on a prediction of the user's average channel gain. Using this prediction enables our scheme to plan its resource allocation ahead. This enables to compensate for upcoming channel outages (e.g., when a user drives through a tunnel) by pre-loading the video buffer in advance. Although this idea of anticipation has been applied for software interfaces \cite{nadin00:anticipatory} and cognitive radios \cite{tadrous11:proactive_ra}, it has not been proposed for media streaming so far.

Further benefits of our work are its generality and low computational complexity. Being based on bit rates, our buffer model captures arbitrary video and audio codecs, while being independent of subjective quality metrics. This level of tractability is not provided by the Utility-based formulation in \cite{huang:08}. Finally, our buffer model enables to incorporate the dynamics of streaming media traffic into a Linear Program. Such low complexity is not provided by the combinatorial approach in \cite{draexler13:span}.
 
\section{Anticipatory Buffer Control\\ and Resource Allocation}\label{contribution}
 
The proposed approach is based on the fundamental assumption that the \emph{average} channel gain is predicted and known for the video user over the next $T$ time slots which is referred to as the \emph{look-ahead window}. The optimization problem is formulated in the context of orthogonal frequency division multiple access (OFDMA) scheme as in Long Term Evolution (LTE) standard. There are $N$ subchannels - physical resource blocks (PRBs) in LTE - available in the system each with a bandwidth of $B$. We assume a cellular communication system with the BS at the centre of each cell. A frequency reuse of 1 is assumed among the cells resulting in an average interference power within each cell. The video user requires an average data rate of $R$ bits/s. For convenience, the rate achieved on a subchannel is assumed to be given by the Shannon capacity. The gap function, however, is used to modify the effective signal-to-interference-plus-noise ratio (SINR) to account for the bit error rate in practical modulations.

\subsection{Play-out Buffer Model}
\begin{figure}
\center
\includegraphics[scale = 0.5]{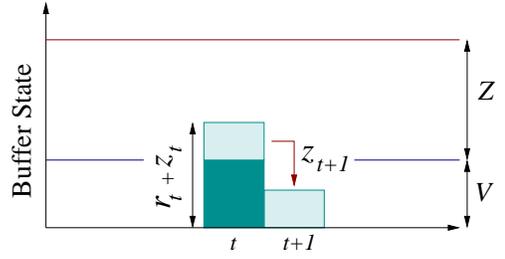}
\caption{Buffer model: at each time slot, the allocated and the carry-over bits are adjusted to have a sum no less than the required video bit rate to avoid outage. The carry-over is limited by the maximum buffer size while the allocated rate is constrained by the link budget.}
\label{buffer}
\end{figure}

The proposed buffer model is illustrated in Fig.~\ref{buffer}. The play-out time $T$ is divided into time slots indexed by $t$ each with a duration of $T_{d}$ seconds. The data rate $r_{t}$ denotes the allocated bits per time slot and $z_{t}$ represents the bits already in the buffer carried over from the previous time slot to the current one.   
The video user requires an average of $V$\footnote{Each time slot can be of arbitrary duration with the required video bit per time slot proportioned accordingly.} bits per time slot in the play-out buffer to smoothly play the video. The total number of bits in the buffer at any time slot is therefore $r_{t} + z_{t}$. Outage occurs if this number is less than $V$; otherwise, $V$ bits are played and the excess, $z_{t+1} = [(r_{t} + z_{t}) - V]^{+}$, will be carried over to the next time slot. $Z$ is the maximum number of carried over bits which indirectly limits the length of the buffered video. The buffer has only $r_{1}$ bits in the first time slot since $z_{1} = 0$. In this model, the outage is avoided if:
\begin{equation}
\begin{cases}
	r_{1} - z_{2} = V, & \\ 
	r_{t} + z_{t} - z_{t+1} = V, & t = 2, 3, . . . , T-1 \\
	r_{T} + z_{T} = V \\   
	\end{cases} \\ \\
\end{equation}
which in the matrix form can be written as: 
\begin{equation}
\textbf{A}\textbf{x} - \textbf{V} = \textbf{0},
\end{equation}
where $\textbf{x}= [r_{1},  r_{2},  \dots r_{T}, z_{2},  \dots z_{T}]^\top$, $\textbf{A}$ is the $T \times (2T-1)$ matrix given by: 
\begin{equation}
\textbf{A} = \begin{bmatrix} \textbf{I} & \textbf{B}  \end{bmatrix}, \nonumber
\end{equation}
$\textbf{I}$ is the identity matrix of the order $T$ and $\textbf{B}$ is given by:
\begin{equation}
\textbf{B} = \begin{bmatrix} -1 	& 0 	& 0 	& \dots 	& 0 \nonumber \\ 
			 1  	& -1	& 0	& \dots 	& 0 \nonumber\\
			 0	&  1	& -1	& \dots 	& 0 \nonumber\\
			 \vdots & \vdots & \vdots & \vdots & \vdots \nonumber\\
			 0 	& 0	& 0	& \dots 	& 1\nonumber\\
			 \end{bmatrix}. \nonumber
\end{equation}
Finally, \textbf{V} is the video bit rate vector of size $T$ elements of which are $V$. 

\subsection{Optimization Problem}
Based on this model, we formulate an optimization problem with a two-fold purpose: 
\begin{enumerate}
\item To incorporate the future channel condition of the video user in the optimization problem.
\item To optimize the video user's requested rates $\{r_{t}\}_{t=1}^{T}$ to avoid outage by buffering the video while imposing the minimum load on the system.  
\end{enumerate}

Generalizing the model for multiple video users, let $\textbf{x}_{k} = [r_{k,1},  r_{k,2},  \dots r_{k,T}, z_{k,2},  \dots z_{k,T}]^\top$ be the optimization vector for video user $k$, with the buffer model as described above. The optimization problem with the objective of imposing the minimum load on the system is formulated as:
\begin{equation}\label{OverTime}
\begin{array}{ll}
\displaystyle \; \; \; \; \; \; \;  \min_{\bf{x_{k}}}\sum_{k}\sum_{t=1}^{T}w_{k,t}  \\ 
\mbox{subject to:}  \\ \\
\mbox{C1:}\;\;\; \;  \displaystyle \textbf{A}\textbf{x}_{k} - \textbf{V}_{k} = \textbf{0} \hspace{0.5in}    \forall k,  \\ 
\mbox{C2:}\;\; \; \;  0 \leq \sum_{k}w_{k,t}\leq N \hspace{0.3in}       t = 1,2,3, \dots T,  \\ 
\mbox{C3:}\; \; \; \;  0 \leq z_{k,t} \leq Z_{k} \hspace{0.53in}       t = 2,3, \dots T \; \; \forall k, \\
\mbox{C4:}\; \; \; \;  r_{k,t} = w_{k,t}T_{d}B\log_{2}\left(1 + \frac{P |\hat{h}_{k,t}|^2}{N (\sigma^{2} + I)} \right), \\
\end{array}
\end{equation}
where C1 is the no-outage constraint for each video user with $V_{k}$ video bit rate requirement. $w_{k,t}$ is the number (can be a fraction) of subchannels required for user $k$ at time $t$. C2 is the constraint on the total available number of subchannels at each time slot. C3 ensures that the number of carry-over bits does not exceed $Z_{k}$ which limits the buffer size for each user. Finally, C4 incorporates the channel prediction in the optimization problem. $\hat{h}_{k,t}$ denotes the estimated \emph{average} channel gain for video user $k$ at time slot $t$. $P$ is the total transmit power and $\sigma^{2}$ and $B$ are the noise power and the bandwidth of each subchannel, respectively. $I$ is the average intercell interference power. 

The problem formulated in (\ref{OverTime}) minimizes the amount of spectrum required to meet each video user's bit rate requirement considering predicted average channel condition in the upcoming time slots. This amount of spectrum, $\{w_{k,t}\}_{t=1}^{T}$, is directly related to the rate to be requested, $\{r_{k,t}\}_{t=1}^{T}$, at each time slot to avoid outage. 

\begin{figure}
\center
\includegraphics[width=\columnwidth]{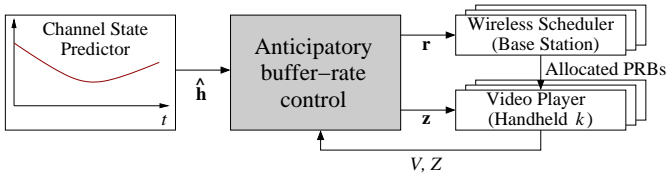} 
\caption{System concept for anticipatory buffer-rate control.}
\label{embodiment}
\end{figure}

Figure~\ref{embodiment} illustrates how our approach would be integrated into a wireless communication system. The set of estimated required rates provided by our scheme would be passed to a scheduler as new rate constraints for each user. While trying to fulfill these rates for the video users, the scheduler optimizes the frequency-time allocation to all its users considering \emph{instantaneous} channel state. 
\section{Performance Study}\label{results}
In this section, we evaluate the performance of the proposed anticipative buffer and resource allocation. The numerical results for a single video user are presented as the proof of concept. We then show the performance of the scheme when multiple video users request streaming video.  

\subsection{Scenario and Parameters}
\begin{figure}
\center
\includegraphics[scale = 0.7]{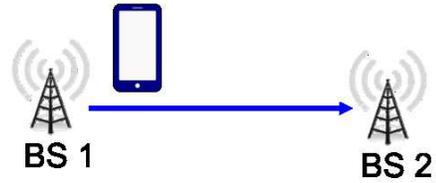} 
\caption{Essential scenario to illustrate buffer-rate allocation: two cells with media streaming users moving from the left into the right cell.}
\label{highway}
\end{figure}

\begin{figure}
\center
\includegraphics[scale = 0.35]{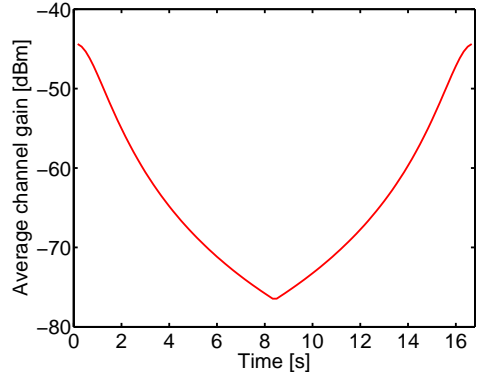}
\caption{Average channel gain for the two-cell scenario.}
\label{channel}
\end{figure}

A two-cell scenario is illustrated in Fig.~\ref{highway}. The video user moves away from BS1 and is admitted by BS2 experiencing first decreasing and then increasing channel gain. The look-ahead window (assumed to be smaller than the length of the video) is 16 seconds over which the user leaves cell 1 and is admitted by cell 2 at the speed of 30 meters per second roughly the moving speed of the highway. This simulation setup has been chosen to capture the essence of the proposed scheme based on channel prediction. It can also emulate a scenario where the user is traversing an overloaded cell with the possibility of good reception in the future based on its trajectory. The corresponding average channel gain is shown in Fig.~\ref{channel}. 

The path loss between the BS and the user accounts for outdoor propagation given by \cite[Table A.2.1.1.2-3]{3GPP}: PL =  $128.1 +  37.6 \log_{10}(d) + L_{s}$ where $d$ is the distance between the serving BS and the user in kilometers. Without loss of generality, we assume that the user connects to the BS with the highest average received power. $L_{s}$ accounts for shadowing and is modelled by log-normal random variable with standard deviation of 10dB. 
Table I lists the parameters used in the simulations, unless otherwise stated. 
\begin{table}
	\begin{center}
	\caption{Parameter Values}
	\begin{tabular}{| c | c |}
	\hline
	Parameters & Value \\
	\hline \hline
	Channel bandwidth &  10 MHz \\
	Resource Block $B$ & 180 kHz \\
	Total Number of PRBs & 50 \\
	\hline
	Transmit Power & 46 dBm \\
	Antenna gain & 0 dB \\
	Antenna & 1 x 1\\
	\hline
	Noise PSD & -174 dBm/Hz \\
	NF & 9 dB\\                           
	I & -149 dBm/Hz \\
	\hline
	Cell diameter & 500 m \\
 	Minimum distance from BS & 35 m \\
	\hline
	Video bit rate & 1.5 Mbits/s\\
	\hline
	\end{tabular}
	\end{center}
\end{table}

\subsection{Proof of Concept for a Single Video User}
\begin{figure}
\center 
\includegraphics[scale = 0.45]{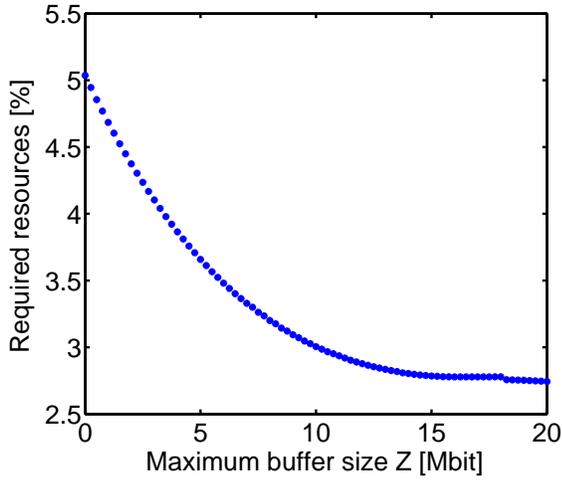}
\caption{Total required spectrum versus buffer size.}
\label{PRBvsBuffer}
\end{figure}

The positive effect of buffering can be clearly seen in Fig.~\ref{PRBvsBuffer}. The figure shows the total amount of spectrum required by the video user for the whole duration of $T$ time slots in the same setup but versus different buffer size constraints $Z$. The required spectrum is normalized by the total number of PRBs available in the system. In the worst case with zero buffer size, the user can not pre-load. Therefore, It has to receive the required video bit rate at each time slot even in poor channel condition hence consuming the highest total amount of spectrum. As the buffer size increases, the user can pre-load more when closer to BS1; hence a smaller number of PRBs is needed. This however, will saturate at some point where the user pre-loads most of the video and plays later. Although desirable from the spectral efficiency point of view, it seems undesirable due to user's unpredictable behaviour as discussed before. In other words, it will be a waste of resources, if the user switches to another video without watching the pre-loaded one. Hence, it is preferable to limit the maximum buffer size.  

\begin{figure}
\center
\includegraphics[scale = 0.55]{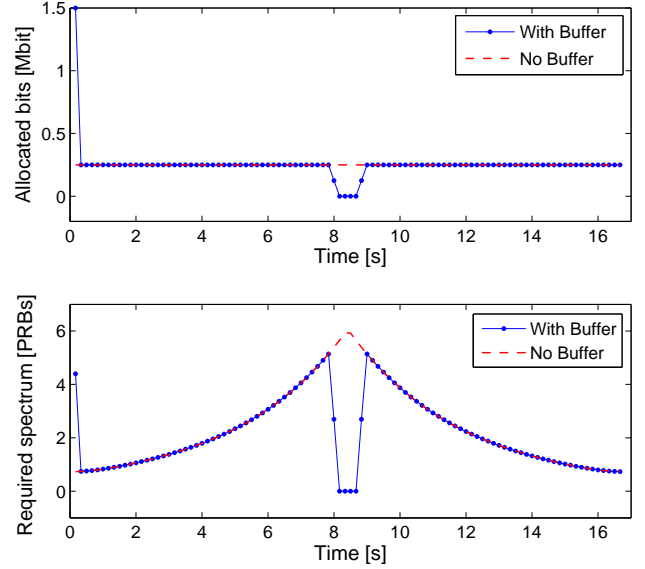}
\caption{Allocated bits $r_{t}$ and spectrum $w_{t}$ to the video user versus time. $T_{d} = 0.167 s$ and $V = 250$ Kbits per time slot. $Z = 5V$ for the pre-buffering user.}
\label{RateSpecVsTime}
\end{figure}

Figure~\ref{RateSpecVsTime} shows the allocated bits and the corresponding required number of PRBs over time as the user moves. $T_{d} = 0.167 s$ and $V = 250$ Kbits per time slot corresponding to $R = 1.5$ Mbits/s video rate. Two cases are plotted: 1) non-zero buffer size and 2) zero buffer size. The latter essentially captures the performance of a user with no possibility of pre-buffering either due to video time constraints or no knowledge of future channel condition. As expected, the video is pre-loaded when the user is close to BS1 to prevent the outage at the cell edge with the worst channel gain. This behaviour minimizes the required spectrum given a maximum buffer size.

\begin{figure}[h]
\center
\includegraphics[scale = 0.55]{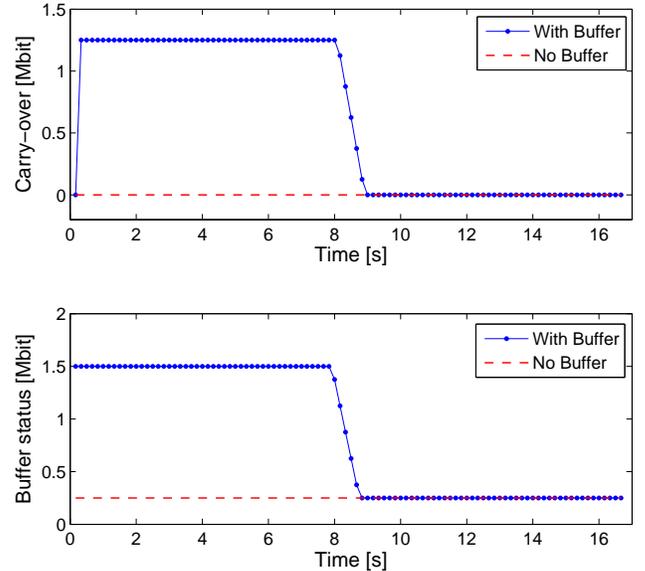}
\caption{Buffer status and Carry-over bits $z_{t}$ versus time. $T_{d} = 0.167 s$ and $V = 250$ Kbits per time slot. $Z = 5V$ for the pre-buffering user.}
\label{BuffCarryVsTime}
\end{figure}

The buffer status and the excess number of bits carried over at each time slot are shown in Fig.~\ref{BuffCarryVsTime}. Four phases can be identified in both figures for the user with anticipatory buffer control: 1) In the first time slot, the user pre-loads the buffer. 2) In the subsequent time slots, it keeps the buffer full while playing the video. The cost of the full buffer increases as the user gets further away from the BS. 3) The increase in the required spectrum and the knowledge of improved future channel gain result in allocating less bits at the cell edge and playing the pre-loaded buffer instead. While the buffer remains full in the previous two phases, it is used up during this phase. 4) Finally, the user is admitted to cell 2, the channel gain increases and with this knowledge, the user only receives $V$ bits per time slot to play the video without any buffering. At this stage, the user does not need to pre-load since a good average channel gain is predicted. As clear in both figures, the main gain is achieved by playing the video during the time the undesired channel condition is experienced. 

\subsection{Performance for Multiple Video Users}
We use the same two-cell scenario with a total number of $K_{v}$ video users requesting service from BS1. The inter arrival time between the users is modelled by an exponential random variable with mean of 0.58 seconds. After the BS1 receives a service request, it estimates the required spectrum as in (\ref{OverTime}). If the system is not overloaded, the user is admitted and the estimated available spectrum over the look-ahead window is updated. In the mean time, the resources are scheduled among the admitted video and already existing users at each time slot. The service request at the required video quality is rejected if the estimated required spectrum is not available.  

\begin{figure}
\center
\includegraphics[scale = 0.45]{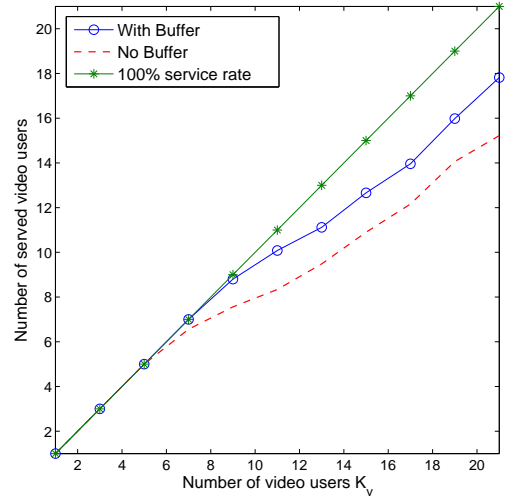}
\caption{Number of served video users versus the total number of service requests $K_{v}$. $R= 1.5$ Mbits/s for all video users and $Z = 5V$. 100$\%$ service rate corresponds to the case where all the users receive their required video bit rate in the time frame under consideration.}
\label{AddUsers}
\end{figure}

Figure~\ref{AddUsers} shows the number of video users that can be supported versus the total number of service requests. Available number of PRBs $N = 15$ to count for the existing load of each cell. Anticipatory buffer control and resource allocation allows for servicing higher number of video users by adjusting the requested rates considering the future load. Such consideration is not possible in a system with no future channel prediction. Hence, although admitted by the system, either the number of video users experiencing no outage decreases or the video user is served with a lower video quality. 

\section{Conclusion}\label{conclusion}
In this paper, we proposed a buffer control and resource allocation scheme for the streaming of non-real time video in cellular networks. 

The proposed formulation has three important benefits. Firstly, it takes advantage of the future user trajectory for a period of time referred to as ``look-ahead" window, reported by the user or predicted and available at the serving BS. This information is translated into the average channel gain and is used in formulating the optimization problem. Such information gives an insight about the possibility of undesired channel condition which in turn mandates the serving BS to increase the allocated data rate for pre-buffering. However, the increase in the data rate is in control; the manner and timing to do so is adjusted consuming the minimum bandwidth for smooth video streaming.

Secondly, our scheme is based on an objective quality measure, i.e., average video bit rate. This makes our formulation general and relevant, while ignoring subjective quality metrics which may be affected by the video content or time of the day. We showed that with the increase in the buffer size for the user with the undesired future channel condition, we can save more resources but at the risk of the user switching the video before watching the pre-loaded one. However, given a maximum buffer size, the scheme optimizes the rate allocation for a smooth video streaming. 

Finally, the proposed scheme is based on linear programming, which makes the implementation easy and fast. In the next step, we will study the trade-off between the error associated with the channel prediction and the achieved gain.
\bibliographystyle{IEEEtran}
\bibliography{CommLett.bib}

\begin{thebibliography}{1}
\providecommand{\url}[1]{#1}
\csname url@samestyle\endcsname
\providecommand{\newblock}{\relax}
\providecommand{\bibinfo}[2]{#2}
\providecommand{\BIBentrySTDinterwordspacing}{\spaceskip=0pt\relax}
\providecommand{\BIBentryALTinterwordstretchfactor}{4}
\providecommand{\BIBentryALTinterwordspacing}{\spaceskip=\fontdimen2\font plus
\BIBentryALTinterwordstretchfactor\fontdimen3\font minus
  \fontdimen4\font\relax}
\providecommand{\BIBforeignlanguage}[2]{{%
\expandafter\ifx\csname l@#1\endcsname\relax
\typeout{** WARNING: IEEEtran.bst: No hyphenation pattern has been}%
\typeout{** loaded for the language `#1'. Using the pattern for}%
\typeout{** the default language instead.}%
\else
\language=\csname l@#1\endcsname
\fi
#2}}
\providecommand{\BIBdecl}{\relax}
\BIBdecl

\bibitem{cisco}
{Cisco}, ``Cisco visual networking index: Global mobile data traffic forecast
  update, 2011-2016,'' \emph{White paper}, Feb. 2012.

\bibitem{pantos11:hls_draft_ietf}
R.~Pantos and W.~May, ``{HTTP} live streaming,'' {IETF}, Informational
  Internet-Draft draft-pantos-http-live-streaming-07, Sep. 2011.

\bibitem{shih-fu:05}
S.-F. Chang and A.~Vetro, ``{V}ideo adaptation: Concepts, technologies and open
  issues,'' \emph{in Proc. of the IEEE}, vol.~93, no.~1, pp. 148--158, Jan.
  2005.

\bibitem{reisslein97:vbr_prefetching}
M.~Reisslein and K.~Ross, ``A join-the-shortest-queue prefetching protocol for
  {VBR} video on demand,'' in \emph{Proc. of International Conference on
  Network Protocols}, Oct. 1997, pp. 63--72.

\bibitem{draexler13:span}
M.~Draexler and H.~Karl, ``Cross-layer scheduling for multi-quality video
  streaming in cellular wireless networks,'' in \emph{Proc. of Int. Wireless
  Communications \& Mobile Computing Conference {(IWCMC)}}, Jul. 2013, to
  appear.

\bibitem{huang:08}
J.~Huang, Z.~Li, M.~Chiang, and A.~K. Katsaggelos, ``{J}oint source adaptation
  and resource allocation for multi-user wireless video streaming,'' \emph{IEEE
  Trans. Circuits and Syst. Video Technol.}, vol.~18, no.~5, pp. 582--595, May
  2008.

\bibitem{nadin00:anticipatory}
M.~Nadin, ``Anticipatory computing,'' \emph{Ubiquity}, vol. 2000, Dec. 2000.

\bibitem{tadrous11:proactive_ra}
\BIBentryALTinterwordspacing
J.~Tadrous, A.~Eryilmaz, and H.~E. Gamal, ``Proactive resource allocation:
  Harnessing the diversity and multicast gains,'' \emph{Arxiv Computing
  Research Repository}, vol. abs/1110.4703, 2011. [Online]. Available:
  \url{http://arxiv.org/abs/1110.4703}
\BIBentrySTDinterwordspacing

\bibitem{3GPP}
{3GPP}, ``{F}urther advancements for {E}-{UTRA} physical layer aspects,''
  {3GPP}, Tech. Rep. TR 36.814 V9.0.0, Mar. 2010.

\end{thebibliography}
\end{document}